# Capacity of Ultra Wide Band Wireless Ad Hoc Networks

Rohit Negi, *Member, IEEE* and Arjunan Rajeswaran, *Student Member, IEEE*

*Abstract*— Throughput capacity is a critical parameter for the design and evaluation of ad-hoc wireless networks. Consider $n$ identical randomly located nodes, on a unit area, forming an ad-hoc wireless network. Assuming a fixed per node transmission capability of $T$ bits per second at a fixed range, it has been shown that the uniform throughput capacity per node $r(n)$ is $\Theta\left(\frac{T}{\sqrt{n \log n}}\right)$, a decreasing function of node density $n$.

However an alternate communication model may also be considered, with each node constrained to a maximum transmit power $P_0$ and capable of utilizing $W$ Hz of bandwidth. Under the limiting case $W \to \infty$, such as in Ultra Wide Band (UWB) networks, the uniform throughput per node is $O\left((n \log n)^{\frac{\alpha-1}{2}}\right)$ (upper bound) and $\Omega(\frac{n^{(\alpha-1)/2}}{(\log n)^{(\alpha+1)/2}})$ (achievable lower bound).

These bounds demonstrate that throughput increases with node density $n$, in contrast to previously published results. This is the result of the large bandwidth, and the assumed power and rate adaptation, which alleviate interference. Thus, the effect of physical layer properties on the capacity of ad hoc wireless networks is demonstrated. Further, the promise of UWB as a physical layer technology for ad-hoc networks is justified.

*Index Terms*— Information theory, ad-hoc network, network capacity, ultra wide band.

## I. INTRODUCTION

It has been shown [1] that the uniform throughput capacity per node of an ad hoc wireless network with $n$ nodes *decreases* with $n$ as $\Theta\left(\frac{T}{\sqrt{n \log n}}\right)$, under certain physical layer assumptions. However, in this paper, it is shown that under an Ultra-Wide Band (UWB) [2] communication model (large bandwidth, limited power), the uniform throughput capacity per node *increases* as $\tilde{\Theta}(n^{(\alpha-1)/2})$, where $\alpha \geq 1$ is the distance loss exponent. (The notation used is elaborated later in the paper.) Thus, this paper shows that, the design of and assumptions about, the physical layer of an ad hoc network can dramatically affect the network capacity.

Wireless communication networks consist of nodes that communicate with each other over a wireless channel. Some wireless networks have a wired infrastructure of controllers, with nodes connected to a controller over a wireless link. Other networks, such as ad-hoc networks [3], are all wireless. Some salient features of ad hoc networks are speedy deployment, low cost and low maintenance. These features lend towards applications such as sensor networks or military systems. The lack of centralized control is also advantageous for short-lived commercial networks.

Ad-hoc network design is burdened with the issues of medium access (scheduling) at the link layer and relaying of data packets (routing) at the network layer. The wireless medium or channel is a resource that must be shared amongst various nodes, a functionality achieved by Medium Access Control (MAC). The broadcast nature of the wireless medium and the decentralized nature of ad-hoc networks makes this scheduling problem very different from that in infrastructure networks. Interference mitigating techniques and distributed protocols have been considered in this regard [4], [5], [6]. Routing is the functionality of transporting data from the source to the destination across a sequence of links. In ad-hoc networks, routing faces a number of challenges, including the variability in topology due to the unreliable wireless link [7] and node mobility. These issues differ significantly from their counterparts in both cellular networks and wired networks (such as the Internet), and have been extensively studied [8], [9], [10]. Traditionally, rate and power adaptation, scheduling and routing are treated independently, as separate functions of the physical, link and network layers, respectively. In wireless ad-hoc networks, however, all nodes use the shared wireless medium resulting in a *strong interdependency between protocol layers*. Therefore, in wireless ad-hoc networks, power adaptation, scheduling and routing require a joint consideration to optimize network performance [11]. In the general case, this joint formulation is a hard non-convex problem [11]. Thus, towards finding appropriate algorithmic solutions it is necessary to develop intuition based on information theoretic analysis.

Recently, there has been significant interest in computing the capacity of ad-hoc networks [1], [12], [13], [14]. The assumed *network model* is a static ad-hoc wireless network, where $n$ identical nodes on a unit area communicate over a wireless channel, with possible cooperation, to relay traffic. The assumed *physical model* is one where each link operates at a fixed data rate, utilizing a finite bandwidth and large power (i.e., large signal-to-noise ratio). Under these assumptions, it was shown in [1], that the per node throughput capacity is $\Theta\left(\frac{T}{\sqrt{n \log n}}\right)$ a decreasing function of the number of nodes $n$. This pessimistic result, has motivated several works that have presented different results, under varied assumptions. Variations on the network model, towards improved capacity results, have been explored. Reference [12] exploited node mobility, to

This work was presented in part at INFOCOM 2005

This work was supported in part by the National Science Foundation under Career award 0347455.

R. Negi is an Associate Professor at the department of Electrical and Computer Engineering, Carnegie Mellon University

A. Rajeswaran is pursuing his Ph.D. at the department of Electrical and Computer Engineering, Carnegie Mellon University

demonstrate that a constant ($\Theta(1)$) per node throughput could be achieved in a mobile network at the cost of a large incurred delay. The issue of delay was addressed in [14], where the assumed mobility models were analyzed to provide bounds on the delay experienced by the packets. A comprehensive result was reported in [15], showing the optimal delay-throughput tradeoff for both static and mobile networks. Hybrid network models have been considered, where the inclusion of base stations (wired infrastructure support), cluster heads (wireless support) or a different number of source and destination nodes introduces an asymmetry into the network. This asymmetry is exploited and detailed analysis in [13] (and references therein) provides improved capacity bounds for these partially ad-hoc networks. Other variations at the network level include altered traffic patterns (e.g. single destination, relays). Simulation-based results on the capacity of small ad-hoc networks [16] confirm the theoretical results in [1], to a limited extent.

Recently, physical layer characteristics and techniques have also been considered in the analysis of capacity. A constant order throughput increase, by the utilization of directional antennas, was demonstrated in [17]. Fading was analyzed in [18] and shown to cause a marginal (in order) loss in capacity. Sophisticated information theoretic ideas such as cooperative relaying and interference cancellation utilized in [19], resulted in bounds similar to [1], providing no gain. Thus, a direct application of physical layer techniques provides little gain.

Currently deployed commercial wireless networks are built using either narrowband (e.g., cellular GSM) or wideband (e.g., 3G, 802.11) links. However, the demand for higher data rates at short distances has created a market for *ultra-wide band* links. UWB radios were designed for covert military applications, as a spread spectrum technology that used a large amount of bandwidth at extremely low powers, thus differentiating them from narrowband and wideband radios. However, recent changes in U.S. federal regulations have opened up UWB for commercial applications. Thus, there are currently intense research and development efforts underway, to design and standardize commercial UWB radios [20], [21], [22]. For example, the UWB based IEEE 802.15.3.a standard [23] is expected to support 100 to 500 Mbps depending on the link distance. Further, UWB radios will be inexpensive and low power, making them ideal for ad hoc wireless applications. Commercial deployment of UWB networks is expected to occur in the near future [24].

Since UWB radios possess properties (extremely large bandwidths and low power) that are drastically different from existing commercial radios (finite bandwidth and large power), the question arises whether the existing results on ad hoc network capacity are applicable to UWB networks. Note that a system with infinite bandwidth does not imply an arbitrarily large link capacity, because of the finite power constraint (See Section IV). Specifically, in this paper, an UWB communication model is assumed, where each link operates over a relatively large bandwidth ($W$) and with a constraint $P_0$ on the maximum power of transmission. The ambient Gaussian noise power spectral density is $N_0$ and the signal power loss, with distance $d$, is $1/d^\alpha$. Here $\alpha \geq 1$, is the distance-loss exponent [28]. Nodes are assumed to remain static. This communication model is analyzed to address the question of network capacity. Upper and lower bounds are derived, demonstrating that the network capacity is an *increasing* function of node density $n$, as $\tilde{\Theta}(n^{(\alpha-1)/2})$, where $\tilde{\Theta}$ stands for *soft order*[1], in contrast to [1], which shows a decreasing function for capacity. (As argued in Section III, this result is equally valid for low power, *low data rate* sensor networks, although they may not use a large bandwidth.) We demonstrated this dramatic effect of the (UWB) physical layer on network capacity first, in [25]. Subsequently, our result has been extended in [26] to tighten the lower and upper bounds to the same order, i.e., the uniform throughput capacity is $\Theta(n^{(\alpha-1)/2})$. This was achieved by changing the assumed network model to one with a random Poisson number of nodes and applying the percolation theory developed in [27] to drop the $\log n$ terms in the bounds. The UWB network capacity result [25] has also been used as a benchmark to validate cross-layer optimization results for UWB networks, such as in [11].

The rest of the paper is organized as follows. In Section II the required background is reviewed. A short intuitive review of the relevant assumptions and proofs from [1] is presented. The concept of uniform convergence is discussed. Section III presents the UWB communication model assumed, motivating the need for an analysis that is different from existing literature. In Section IV, important characteristics of the alternate communication model are presented and are used to show that the optimal MAC is Code Division Multiple Access (CDMA). Section V derives an upper bound on the uniform throughput capacity. This is achieved by analyzing an optimal routing scheme, under a relaxed power constraint. In Section VI, a lower bound is derived by applying some results of [1] to the new communication model. It is thus shown that the uniform throughput capacity $r(n)$ is $\tilde{\Theta}(n^{(\alpha-1)/2})$, in contrast to [1] where $r(n)$ is $\Theta(\frac{1}{\sqrt{n \log n}})$. Bandwidth scaling is addressed in Section VII as a guideline for practical implementations. Finally, conclusions are presented in Section VIII.

## II. BACKGROUND

The objective of this paper is to demonstrate the effect of the ( UWB ) physical layer on ad hoc network capacity. This requires contrasting the assumptions and results of [1] with our results. Therefore, for the sake of completeness, this section reviews the assumptions and methods used in [1], that are most relevant for such a comparison. Readers familiar with [1] may skip this section.

### A. Network Model

The assumed network model in [1], relevant to this paper, is one of Random Networks, where the $n$ static nodes on a unit area are i.i.d. (independent and identically distributed) and distributed uniformly. To avoid edge effects, this unit area is considered to be on the surface of a sphere, $S^2$. The $n$ nodes communicate over a wireless channel, with possible cooperation, to relay traffic. Each of the $n$ nodes has an independent randomly chosen destination (chosen as the node closest to a random point, i.e. uniformly and independently distributed).

---

[1] Here $O, \Omega$ and $\Theta$ are the standard order bounds. The soft order $\tilde{\Theta}$ is the same as a $\Theta$ bound with the powers of $\log n$ neglected.

### B. Performance Metric

All nodes require to send traffic at a rate of $r(n)$ bits per second to their corresponding destinations. A uniform throughput $r(n)$ is feasible if there exists a scheduling and relaying scheme by which every source-destination pair can communicate at a time-average of $r(n)$ bits per second. The maximum feasible uniform throughput is the uniform throughput capacity, and is the metric of choice. The motivation for choosing this metric is a sense of fairness, since all nodes are assumed to be homogeneous in their capabilities and requirements.

Since the underlying network is random, so is the capacity. The aim is to bound the random capacity by functions of $n$. Thus, bounds are shown that hold with *w.h.p.* (the abbreviation *w.h.p.* represents 'with high probability', i.e., with probability approaching 1 as $n \to \infty$). Specifically, the uniform throughput capacity $r(n)$ is said to be of order $\Theta(f(n))$ if there exists deterministic constants $k_1 > k_0 > 0$ s.t.

$$\lim_{n \to \infty} \text{Prob}(\ r(n) = k_0 f(n) \text{ is feasible }) = 1$$
$$\lim_{n \to \infty} \text{Prob}(\ r(n) = k_1 f(n) \text{ is feasible }) < 1. \quad (1)$$

($k_i, K_i, c_i, C_i$ will be used to denote constants, with respect to $n$.)

### C. Communication Model:

In [1], it was assumed that each node can transmit at a rate of $T$ bits/s. The homogeneous nodes share a common range (equivalently, power) of transmission $tr(n)$. This simplistic communication model assumes that all operating links transport data at a constant rate. The transmission by a legitimate transmitter $X_i$ to its intended receiver $X_j$ is successful, if their distances are related as ,

$$|X_k - X_j| \geq (1 + \Delta) |X_i - X_j| \quad (2)$$

for every other $X_k$ transmitting simultaneously. This interference criterion models a protocol, which specifies a guard zone $\Delta$, around the receiver where no other node may transmit, and is termed as the 'protocol model'. A second model, more directly related to physical layer design, works with the Signal to Interference Noise Ratio at the receiver, *with the assumption of arbitrarily large power*. (Note that the assumption of arbitrarily large power is in sharp contrast to the UWB model assumed in this paper.) The order results remain the same under both the 'protocol and physical' models. Note that both these models for the physical layer do not allow for rate and power adaptation. For simplicity this review assumes the protocol model.

### D. Network throughput

With these assumptions it is proven in [1] that the capacity of random ad-hoc networks

$$r(n) \text{ is } \Theta(\frac{T}{\sqrt{n \log n}}). \quad (3)$$

Thus the throughput per node decreases with increasing node density. The essential reason for this capacity decrease is the requirement for all nodes to share the wireless channel locally. This may be demonstrated by a contrast between the MAC and routing requirements. The mean source-destination distance is assumed to be $L$. The mean number of hops taken by packets is $\frac{L}{tr(n)}$. The total traffic generated by all nodes due to the multi-hop relaying (routing) is $n\ r(n)\ \frac{L}{tr(n)}$ bits/s. This traffic is required to be served by all nodes. However, the capacity of each node is reduced by interference (MAC), since nodes close to a receiver cannot transmit simultaneously. The interference radius is proportional to the transmission radius (2). Since the number of interfering nodes is proportional to the interference area (uniform distribution), the capacity loss is quadratic in the transmission radius $tr(n)$. Thus, the available capacity reduces from $nT$ to $\frac{T}{tr(n)^2}$. The tradeoff between the routing requirement and the MAC restriction yields that the capacity of the network

$$r(n) \leq k_2 \frac{T}{tr(n)\ L\ n}\ . \quad (4)$$

A lower limit on $tr$, due to the requirement of network connectivity, has been derived as $tr(n) \geq \sqrt{\frac{\log n}{\pi n}}$. The lower limit on $tr(n)$ ensures that no node in the network is isolated *w.h.p.*. This concept of connectivity in a network arises due to the assumed physical layer model, that requires a *fixed rate* of transmission and hence a minimum distance to allow for a link. The lack of isolation is required, since the performance metric is one of *uniform* throughput capacity, which would be zero if there was even a single starved node. The application of this limit to (4) results in the capacity upper bound as

$$r(n) \leq k_2' \frac{T}{\sqrt{n \log n}}\ . \quad (5)$$

### E. Lower bound

To provide a capacity lower bound (feasibility), specific MAC and routing schemes are chosen. These schemes achieve the same order of capacity as the upper bound. The lower and upper bound prove the order bound (3) on $r(n)$.

To elaborate, it is required to specify schemes for both the MAC and the routing on this random network. Such a specification requires imposing some structure on the random network. Motivated by cellular architectures, a tessellation (covering by 'cells') of the unit area is considered. Regularity in the tessellation (in the properties of every cell) necessitates a regular cell shape. However, since the network is random, some deviation from a regular cell shape should be allowed to ensure that the tessellation may be made as fine as required. **A Voronoi tessellation $\mathcal{V}_n$:** of the surface $S^2$ is a tessellation that has the desired properties. The tessellation (such as in Figure 1) has the following properties :

1) Every Voronoi cell $V$ contains a disk of area $\frac{100 \log n}{n}$ and corresponding radius $\rho(n)$.
2) Every Voronoi cell is fully contained within a circle of radius $2\rho(n)$.

The existence of such a tessellation was proved in [1]. The size of each cell, relative to the number of nodes, is important. It is required that every cell contain at-least one node, to ensure that the routing scheme (described below) is feasible. It is this uniformity restriction that results in the choice of cell size. The resulting 'cellular-like' architecture imparts some notion of regularity on the random network.

**MAC and available capacity:** With the Voronoi tessellation, a MAC is defined that achieves a scheduling between the cells. The MAC ensures that transmissions from a cell do not interfere with transmissions in simultaneously transmitting cells. The radius of transmission is chosen to be $tr(n) = 8\rho(n)$ to allow for direct transmissions between adjacent cells (cells sharing a point) and within a cell (Figure 1). Cells containing nodes within a distance of $(2 + \Delta)tr(n)$ are interfering cells, since a node in one cell may interfere with the transmission in the interfering cell (2). The distance between two nodes in interfering cells is upper bounded. Also, the area of each interfering cell is lower bounded, by the tessellation properties. The ratio of the maximum interference area and the minimum area of each cell is a constant, $k_3$. Thus for every cell in the tessellation the total number of interfering cells may be upper bounded by a constant $k_3$ which depends only on $\Delta$ (the parameter of the interference model). Consequently the graph defined by interference amongst cells, has a bounded degree of $k_3$.

The chosen MAC is a schedule of $1 + k_3$ slots, in which each cell is assigned one slot to transmit. This is possible since a graph with degree not greater than $k_3$ may be colored by $1 + k_3$ colors [29]. Thus, the 'cellular-like' architecture is utilized to achieve a simple slotted MAC amongst the cells.

Therefore each cell has an available capacity of

$$\text{available capacity} = \frac{T}{1 + k_3} \text{ bits/s.} \quad (6)$$

**Routing:** every source destination pair may be connected by a straight line segment (segment of a great circle on $S^2$), as in Figure 1. The packet routing scheme employed is as follows. Packets originating from a source are relayed from the cell containing the source to the cell containing the destination in a sequence of hops. In each hop, the packet is transferred from one cell to another, in the order in which cells intersect the straight line segment connecting the source and destination. A node (head/relay node) is chosen randomly in each cell to relay all traffic. Within a cell all sources send traffic to the head node and destinations receive traffic from the head node. This choice of routing is independent of the MAC and hence the two are analyzed separately. This separation of the MAC and routing may not necessarily be the optimal strategy (from a maximizing throughput perspective), however here it is required to show some scheme that achieves the same order as the upper bound.

To make relaying of traffic between cells feasible, it is required that every cell in the tessellation $\mathcal{V}_n$ contains at least one node *w.h.p.* (uniformity requirement). A simple union bound of the probabilities that every cell contains at least one node is insufficient, and hence, a more intricate technique is required to provide this uniformity. The appendix reviews Vapnik-Chervonenkis (VC) theory, which provides the required uniform convergence in the analysis of the uniform throughput capacity.

**Routing and traffic to be carried:** The VC theorem (33) obtained from VC theory may be applied to the tessellation of the network. A set of disks (of fixed area) have a VC dimension of three. Each cell in the tessellation contains a disk of radius $\rho(n)$ and is contained in a disk of radius $2\rho(n)$. The application of the VC theorem to the set of disks (contained in the cells) yields

$$\text{Prob}\left(\sup_{V \epsilon \mathcal{V}_n} |N(V) - 100 \log n| \leq 50 \log n\right)$$
$$> 1 - \frac{50 \log n}{n}, \quad (7)$$

where $N(V)$ is the number of nodes in cell $V$. The result (applied to both the sets of disks) implies that *w.h.p.*, the network G is such that *for every Voronoi cell in the tessellation,* the number of nodes per cell obeys $50 \log n \leq N(V) \leq 150 \log n$. This result allows for the viability of the routing scheme, by guaranteeing a node in every cell that can serve as the head node, and justifies the choice of cell size as $\frac{100 \log n}{n}$.

The traffic generated due to this routing scheme is considered. The random sequence of straight-line segments is i.i.d and hence the weak law of large numbers may be applied to the routes which approximate these line-segments. The traffic to be carried by a cell is proportional to the number of straight line segments passing through the cell. The number of routes intersecting every cell maybe bounded *w.h.p.*. Thus, the traffic to be carried by every cell can be upper bounded *w.h.p.* as

$$\sup_{V \epsilon \mathcal{V}_n} \text{(Traffic carried by cell V)}$$
$$\leq k_4 \ r(n)\sqrt{n \log n}. \quad (8)$$

**Bounds:** The lower bound is derived by constraining the traffic to be carried (8), obtained from the routing requirements, to be less than the available capacity (6), obtained from the MAC constraint. Thus for random networks

$$r(n) = k_5 \frac{T}{\sqrt{n \log n}} \quad (9)$$

bits/s is feasible *w.h.p.*. The upper bound, obtained by the requirement for connectivity, also presented the same order (5) and hence $r(n)$ is of order $\Theta(\frac{T}{\sqrt{n \log n}})$.

## III. UWB NETWORKS

Next, the capacity of UWB random ad-hoc networks is considered. The assumed random network model as described in Section II, is the same as in [1]. In contrast to existing literature, the following **UWB Communication Model** is assumed to model the physical layer:

1) Power: Each node is constrained to a maximum transmit power of $P_0$.
2) Bandwidth : The underlying communication system has an arbitrarily large bandwidth ($W \to \infty$).

The key characteristic of such a model is the *low spectral efficiency* (i.e., $\frac{P_0}{N_0 W} \ll 1$), which implies a relatively large bandwidth [28] or a relatively tight power constraint. The results of this paper are applicable to all systems that have a low spectral efficiency. Thus, in particular, the results hold for two practical applications,

1) *UWB systems*, where the bandwidth used is of the order of a few GHz, such as in the IEEE 802.15.3a standard [23]. For such a system, $W \to \infty$, which implies that $\frac{P_0}{N_0 W} \ll 1$. As mentioned in the Introduction, such systems are actively being considered for commercial deployment.
2) *Sensor networks*, with bandwidths of the order of a few MHz or less, but which use very low power devices (to extend battery life). For such a low power system, $P_0 \to 0$, which implies that $\frac{P_0}{N_0 W} \ll 1$. Such networks are being considered in both military as well as commercial applications [30].

As noted in Section I, an ambient Gaussian noise power spectral density of $N_0$ and a signal power loss of $1/d^\alpha$, with distance $d$, is assumed. Here $\alpha \geq 1$, is the distance loss exponent. Shadowing effects are not considered in this model. Capacity-achieving Gaussian channel codes are assumed for each link. Thus, each link is assumed to support a data rate corresponding to the Shannon capacity[2] of that link [31]. It is assumed that each node can transmit and receive simultaneously (although this restriction does not affect the results, as shown later). Also, each node can control its transmit power, as well as adapt its data rate to the link condition [32], [33]. Every node may transmit or receive, and wishes to communicate with a randomly chosen destination (chosen as the node closest to a randomly chosen point).

It has been shown in [1] that $r(n)$, the uniform throughput capacity per node, is a decreasing function of $n$ (3), under a simplistic fixed per-link data rate. In the communication model assumed here, the constraints on power and bandwidth are different from [1], the link capacity explicitly depends on distance, and each link is allowed to adapt its power and rate. Therefore, the results of [1] are not applicable in our case. Thus, in contrast, in this paper it is shown that under the new communication (physical layer) model, the capacity of the ad-hoc network *increases* as a function of the node density $n$!

To demonstrate this result, the characteristics of the communication model are studied. This includes a presentation of the optimal MAC, followed by an analysis of the routing problem that provides the required upper and lower bounds on the uniform throughput capacity.

## IV. OPTIMALITY OF CDMA MAC

The interference problem in the ad hoc network is first addressed. It is shown that the interference perceived by a receiver is bounded *w.h.p.*, and hence, a certain scaling of bandwidth $W$, as a function of $n$, renders the interference negligible. This, implies that under the limiting bandwidth assumption, a 'CDMA MAC scheme' is optimal. i.e., all transmitters transmit at the same time, using the entire bandwidth. Here 'optimal' is used in comparison to time/frequency scheduling schemes as noted subsequently.

[2]The Shannon capacity $r$ for a link with Gaussian noise and interference sources is, $r = W \log(1 + SINR)$, where $SINR$ is the signal-to-interference noise ratio of that link.

### A. Bandwidth Scaling

Let $X_i$ denote the node and its position. Let $P_{ij} \geq 0$ be the transmit power chosen by node $X_i$ to transmit to its chosen receiver $X_j$, over link $X_i \to X_j$. The node power constraint $P_0$ implies that $P_i \triangleq \sum_j P_{ij} \leq P_0$. The wireless medium causes a power loss $g_{ij}$, given by $g_{ij} = \frac{1}{|X_i - X_j|^\alpha}$, where other physical constants like antenna gain have been absorbed into $N_0$. The distance $|X_i - X_j|$ is defined as the length of the segment along the great circle, connecting $X_i$ and $X_j$, on the surface $S^2$. The signal-to-interference noise ratio at the receiver $X_j$ is [31]

$$\text{SINR} = \frac{P_{ij} g_{ij}}{W N_0 + \sum_{k \epsilon I} P_k g_{kj}}, \quad (10)$$

where $I$ is the set of all interfering nodes (the set of all simultaneous transmitters). It is required to bound the interference, so that a certain bandwidth scaling can render the interference negligible with respect to ambient noise. The problem stems from the fact that potentially, a node arbitrarily close to the receiver (i.e., $X_k$ s.t. $|X_k - X_j| \to 0$) could cause arbitrarily large interference. This is, however, a very low probability event. Specifically, let the random variable $d_{min}(G)$ denote the minimum distance (on the surface $S^2$ of the sphere) between pairs of nodes in a specific realization $G$ of the the network. The following lemma shows that $d_{min}(G)$ cannot be very small.

*Lemma 1:* $\text{Prob}\left(d_{min}(G) < \frac{1}{n\sqrt{\log n}}\right) \leq \frac{c_2}{\log n}$.

*Proof:*

$$\text{Prob}\left(d_{min}(G) < \frac{1}{n\sqrt{\log n}}\right)$$
$$= \text{Prob}\left(\bigcup_{i>j} \left(|X_i - X_j| < \frac{1}{n\sqrt{\log n}}\right)\right)$$
$$\leq \sum_{i>j} \text{Prob}\left(|X_i - X_j| < \frac{1}{n\sqrt{\log n}}\right)$$
$$\stackrel{(a)}{\leq} n^2 \frac{\pi}{n^2 \log n}, \quad (11)$$

where (a) arises because the uniformly distributed node $X_j$ has to lie within a disc of radius $\frac{1}{n\sqrt{\log n}}$ centered on $X_i$. Thus w.h.p., $d_{min}(G)$ of network G exceeds $\frac{1}{n\sqrt{\log n}}$. ∎

Noting that $P_k \leq P_0$, $|I| \leq n$ and $g_{ij} \leq (n^2 \log n)^{\frac{\alpha}{2}}$ (from (11)), the total interference can be bounded w.h.p. by $P_0 n (n^2 \log n)^{\frac{\alpha}{2}}$. Thus, setting $W = \Theta(n(n^2 \log n)^{\frac{\alpha}{2}})$ renders the interference negligible with respect to ambient noise. (Section VII discusses a practical bandwidth scaling.)

The above bandwidth scaling ensures that there is no requirement to schedule transmitters, since they cause negligible interference to each other. This bandwidth scaling (which implies that $W \to \infty$, as $n \to \infty$), allows for a CDMA MAC, where all nodes may transmit simultaneously. It is proved below that the CDMA MAC is indeed an optimal MAC scheme for such an ad hoc network.

## B. Optimality of CDMA MAC

The optimality of the CDMA MAC is in the sense that it performs at least as well as any other optimal scheduling scheme, which assigns time slots and frequency bands to various nodes (TDMA/FDMA), as shown below.

Since the bandwidth is arbitrarily large, each link's Shannon capacity $r_{ij}$ is proportional to the received power, as below.

$$r_{ij} = \lim_{W \to \infty} W \log(1 + \frac{P_{ij}g_{ij}}{N_0 W}) = \frac{P_{ij}g_{ij}}{N_0}. \quad (12)$$

Throughout this paper, $\log(\cdot)$ denotes $\log_e(\cdot)$ and capacity is expressed in units of nats [31]. As (12) shows, although the bandwidth is infinite, the link capacity is bounded, due to the finite power constraint, a classical result in communication theory.

Now, assume that there exists a TDMA/FDMA scheduling scheme, which coupled with some routing scheme, achieves the maximum possible uniform throughput. Consider the following generic partition of the allotted bandwidth $W$ and the time frame (normalized to unity), corresponding to this optimal TDMA/FDMA solution; $\{W_k, k = 1, 2, \ldots, K\}$, s.t. $\sum_k W_k = W$, and $\{f_t, t = 1, 2, \ldots, T\}$, s.t. $\sum_t f_t = 1$. The assumed optimal TDMA/FDMA scheduling scheme partitions the total power $P_{ij}$ assigned to link $X_i \to X_j$, as $P_{ij}^{k,t}$ s.t. $\sum_{k,t} f_t P_{ij}^{k,t} = P_{ij}$. Thus, $P_{ij}^{k,t}$ is the power assigned to the link in the $t^{th}$ time slot of length $f_t$, and over the $k^{th}$ frequency band of bandwidth $W_k$. The following theorem shows that a CDMA MAC is indeed optimal.

*Theorem 2:* For each link $X_i \to X_j$, the rate $r_{ij}$ achieved using a CDMA MAC scheme is not less than that achieved using the ptimal TDMA/FDMA scheduling scheme.

*Proof:* Consider a particular link $X_i \to X_j$. Since the rate achieved is upper bounded by the capacity in the absence of interference, the rate achieved on this link by the TDMA/FDMA scheme is bounded as,

$$\begin{aligned} r_{ij}^{TDMA/FDMA} &\leq \sum_{k=1}^{K} \sum_{t=1}^{T} f_t W_k \log(1 + \frac{P_{ij}^{k,t} g_{ij}}{N_0 W_k}) \\ &\stackrel{(a)}{\leq} \sum_{k=1}^{K} \sum_{t=1}^{T} f_t \frac{P_{ij}^{k,t} g_{ij}}{N_0} \\ &= \frac{P_{ij} g_{ij}}{N_0}, \end{aligned} \quad (13)$$

where (a) arises since $x \log(1 + \frac{C}{x}) \leq C$. $P_{ij} \leq P_0$ is the total power assigned to link $X_i \to X_j$. Since $\frac{P_{ij}g_{ij}}{N_0}$ is the rate achieved by the CDMA MAC scheme (12), the theorem is proved. ∎

It is thus proven that the arbitrarily large bandwidth renders interference to be negligible and the resulting CDMA MAC is optimal. It should be noted that making an ad-hoc wireless network interference free, alone, is insufficient to drastically improve its capacity. Utilization of highly complex signal processing to generate arbitrarily narrow directed beams at transmitters and receivers results in the removal of all interference. However, even such an impractical technique was shown [34] to provide a capacity gain (compared to [1]) of only $\Theta(\log^2(n))$. As is demonstrated below the assumed UWB physical model provides a combination of mechanisms, including rendering interference negligible, that results in a drastically improved capacity.

The (optimal) CDMA MAC scheme is assumed in the subsequent sections. Essentially, this results in a clean separation of the MAC and routing problems, i.e., it remains to consider optimal routing of the source-destination pairs, with the links scheduled using the CDMA MAC. Since the bandwidth is arbitrarily large, the key constraint is no longer bandwidth, but rather the power of the nodes. Thus, unlike [1], which analyzed the distribution of bandwidth among the different links, in our case, the distribution of power among the competing links and routes needs to be analyzed.

## V. AN UPPER BOUND ON THROUGHPUT CAPACITY

With a CDMA MAC, the optimal routing consists of finding source-destination routes for all sources, that achieve the uniform throughput capacity. The difficulty here is that, the per-node power constraint results in a coupling between the route selections for different sources. However, interestingly, as opposed to the classical routing problem in wired networks [35], the constraint is not in terms of the capacities of individual links, but rather, in terms of the total power transmitted by each node. An upper bound on throughput capacity is derived in this section, by analyzing such a 'power-constrained routing' problem.

The upper bound (Section II) in [1] was derived by bounding the minimum transmission radius as $tr(n) \geq \sqrt{\frac{\log n}{\pi n}}$. It was shown that violation of the bound on $tr(n)$ would result in an isolated node (all neighbors being beyond $tr(n)$), causing the uniform throughput capacity to be zero. However, in this case, due to link adaptation (12), there is no concept of node isolation, or network disconnectivity, since the link capacity simply decreases with distance, but is always non-zero. Therefore, a more sophisticated method, which can analyze the optimal power-constrained routing in detail, is required to upper bound the throughput capacity.

### A. Traffic Routing

The routing problem is to find a set of routes for each source-destination pair, and to find the power to be allotted to each link along these routes, to maximize the uniform throughput capacity of the network. Under the optimal CDMA MAC, each link's Shannon capacity is proportional to the received power (12). Thus,

$$r_{ij} = \frac{P_{ij}g_{ij}}{N_0} \Rightarrow P_{ij} = r_{ij}N_0|X_i - X_j|^\alpha. \quad (14)$$

The coupling of the various routes (due to the per-node power constraint) complicates the routing analysis. Therefore, towards obtaining an upper bound, the power constraint is *relaxed* from a constraint on each node, to an average (or equivalently, total) power constraint. Thus, assume that

$$\sum_{i=1}^{n} P_i \leq nP_0, \quad P_i \geq 0 \ \forall \ i, \quad (15)$$

instead of $P_i \leq P_0 \ \forall \ i$.

Consider the source node $X_i$ and the set of all possible routes from this source to its final destination (recollect that each source is assumed to have exactly one destination). Note that all links on a specific route must operate at an equal data rate. For, if this were not the case, a redistribution of power amongst the links (while maintaining the total power utilized) would result in a new rate which is at least as large as the previous rate. (Such a redistribution is possible due to the relaxed power constraint (15).) Therefore, each route can be associated with a single rate.

Thus, assume some optimal power distribution amongst the set of all routes for a given source-destination pair, for each pair, that results in the maximum uniform throughput. Can this power distribution be characterized? Consider two routes corresponding to a given source-destination pair, $X_i \rightarrow X_i^K$, as $R_i = [X_i^0 X_i^1 X_i^2 .........X_i^K]$ and $R_i^* = [X_i^0 X_i^{1*} X_i^{2*} .........X_i^{K*}]$ where $X_i^0 = X_i^{0*} = X_i$ is the source and $X_i^K = X_i^{K*}$ is the destination. Let $r_i(n)$ and $r_i^*(n)$ be the rates achieved on these routes (i.e., on every link of each route) respectively. The route $R_i^*$ is defined as the route for which

$$\sum_{k=1}^{K*} |X_i^{k*} - X_i^{k-1*}|^\alpha \quad (16)$$

is the minimum of all possible routes from the source to its destination. i.e., $R_i^*$ is the minimum power route for the chosen source-destination pair. From (14), the total power used on these routes is respectively

$$P(R_i) = r_i(n) N_0 (\sum_{k=1}^{k=K} |X_i^k - X_i^{k-1}|^\alpha),$$
$$P(R_i^*) = r_i^*(n) N_0 (\sum_{k=1}^{k=K*} |X_i^{k*} - X_i^{k-1*}|^\alpha). \quad (17)$$

If the power $P(R_i)$ is shifted from the route $R_i$ to $R_i^*$, by scaling the power of each link $X_i^{k*} \rightarrow X_i^{k+1*}$ of $R_i^*$ by a factor $(1 + \frac{P(R_i)}{P(R_i^*)})$, the relaxed power constraint (15) would still be met, while the achieved rate on $R_i^*$ would be not less than $r_i(n) + r_i^*(n)$. This follows from (17) and from the definition (16) of $R_i^*$. Thus, under the relaxed power constraint (15), it is sufficient for each source-destination pair $i$ to *choose the minimum power route $R_i^*$ to route all its traffic*, so as to maximize its rate. Further, different source-destination pairs make their choice independent of other pairs. Note that such a simplification in routing is not possible with the original per-node power constraint ($P_i \leq P_0 \ \forall \ i$).

Therefore, the exact uniform throughput capacity $r^u(n)$ *under the relaxed power constraint* may be obtained by setting $r_i^*(n) = r^u(n), \ \forall \ i$, and solving $\sum_i P(R_i^*) = nP_0$, where $P(R_i^*)$ is given by (17). This routing scheme will be referred to as 'Minimum Power Routing'. In general, this may not coincide with shortest-path routing.

The uniform throughput capacity *with the per-node power constraint* $r(n)$ satisfies $r(n) \leq r^u(n)$. The objective of this section is to upper bound $r(n)$, which is achieved below by upper bounding $r^u(n)$.

As an aside, the per-node power-constrained routing problem may be posed as a convex optimization problem. The problem is similar to the classical joint optimal routing problem for wired networks [35], but differs in that the constraints are per-node power constraints, rather than per-link capacity constraints. Decentralized algorithms to solve the classical joint optimal routing problem [35], and the resulting practical routing protocols, may point to similar solutions for our power-constrained routing problem. This will be the subject of future investigation.

*B. Maximum number of nodes on a route*

As described above, the 'Minimum Power Routing' scheme chooses the minimum power route $R_i^*$ (16) for each source-destination pair $i$. To bound $r^u(n)$, the maximum number of hops in $R_i^*$ is required. Intuitively, if it were possible for $R_i^*$ to have a large number of short hops, then potentially the throughput capacity can become very large, due to rate adaption (12).

Denote $D_i$ as the distance between the source $X_i$ and its destination $X_i^{K*}$ (measured on $S^2$). By the triangle inequality, the sum of the hop-lengths $L_i$ of path $R_i^*$ is lower bounded by $D_i$ as,

$$L_i \triangleq \sum_{k=1}^{k=K*} |X_i^{k*} - X_i^{k-1*}| \geq |X_i - X_i^{K*}| \triangleq D_i \quad (18)$$

Consider the Voronoi tessellation of the network, as described in Section II. Note that $\rho(n)$ is the radius of the circle with area $\frac{100 \log n}{n}$ on the surface of a sphere $S^2$. Also note that

$$4\rho(n) \leq \sqrt{\frac{3200 \log n}{\pi n}}. \quad (19)$$

This is because a circle of radius $\rho$ on $S^2$ has an area less than $\pi \rho^2$ and more than $\frac{\pi}{2} \rho^2$. The following lemma will be used to bound the number of nodes on $R_i^*$.

*Lemma 3:* The number of Voronoi cells $N_{max}$ that intersect a minimum power route $R_i^*$ is upper bounded by $32 + \frac{16 L_i \sqrt{n}}{10 \sqrt{\pi} \sqrt{\log n}}$.

*Proof:* A particular node $X_i$ is considered, along with its corresponding optimal route $R_i^*$. Define a region $C(R_i^*) \subset S^2$ as follows.

$$Y \epsilon C(R_i^*) \quad \text{iff} \quad \exists \ Z \epsilon R_i^* \ \text{s.t.} \ |Z - Y| \leq 4\rho(n),$$

here $Y$ and $Z$ are points on $S^2$.

$C(R_i^*)$ defines a coverage region around the route such that all cells intersecting the route have to be fully contained within this coverage region. We now bound the area of the coverage region. Corresponding to a route, each link contributes a band (rectangular region) of width $4\rho(n)$ and length $|X_i^{k*} - X_i^{k-1*}|$ to the coverage region. Also, the edge links contribute an additional two semi-circular regions of radius $4\rho(n)$. This is demonstrated in Figure 2. Thus, the total area is bounded as

$$\text{Area}(C(R_i^*)) \leq \frac{3200 \log n}{n} + 2 \sqrt{\frac{3200 \log n}{\pi n}} L_i \quad (20)$$

The minimum area of a Voronoi cell is $\frac{100 \log n}{n}$ (Section II). Since the route can only intersect cells that are completely contained in $C(R_i^*)$, the number of such cells is upper bounded as

$$\begin{aligned} N_{max} &\leq \frac{\text{Area}(C(R_i^*))}{\text{Minimum area of a Voronoi cell}} \\ &\leq 32 + \frac{16 L_i \sqrt{n}}{10 \sqrt{\pi} \sqrt{\log n}} \end{aligned}$$

∎

Using the result (7), with probability exceeding $1 - \frac{50 \log n}{n}$, every cell in the tessellation contains at most $150 \log n$ nodes. Thus, the maximum number of nodes on $R_i^*$ is bounded *w.h.p.* as

$$\begin{aligned} N_{max}^{nodes} &\leq (150 \log n) \times (N_{max}) \\ &\leq (c_1 \log n + c_2 L_i \sqrt{n \log n}) \end{aligned} \quad (21)$$

*C. Upper bound on throughput capacity*

The power $P(R_i^*)$ (see 17) utilized on $R_i^*$ is related to the length of the route and the rate achieved on that route. This relation is obtained as

$$\begin{aligned} P(R_i^*) &= r^u(n) N_0 \sum_{k=1}^{k=K^*} |X_i^{k*} - X_i^{k-1*}|^\alpha \\ &\stackrel{(a)}{\geq} N_{max}^{nodes} r^u(n) N_0 \left(\frac{L_i}{N_{max}^{nodes}}\right)^\alpha \\ &\stackrel{(b)}{\geq} r(n) N_0 \frac{L_i^\alpha}{(c_1 \log n + c_2 L_i \sqrt{n \log n})^{\alpha-1}} \\ &\stackrel{(c)}{\geq} r(n) N_0 f(L_i), \end{aligned} \quad (22)$$

where (a) is because of the convexity of $y^\alpha$ for $\alpha \geq 1$, $K^* \leq N_{max}^{nodes}$ and from (18). (b) is from (21) and because $r^u(n)$ upper bounds $r(n)$. (c) is from defining $f(L_i) \triangleq \frac{L_i^\alpha}{(c_1 \log n + c_2 L_i \sqrt{n \log n})^{\alpha-1}}$.

As a final step in deriving the upper bound, the expected total power required by all the $n$ routes, over the ensemble of graphs $G$, is bounded by the total available power $nP_0$. Therefore, by symmetry, the expected power of each route $P(R_i^*)$ is bounded by $P_0$ as

$$\begin{aligned} P_0 &\geq \mathbf{E} P(R_i^*) \\ &\stackrel{(a)}{\geq} r(n) N_0 \, \mathbf{E} f(D_i) \\ &\stackrel{(b)}{\geq} r(n) N_0 \, \text{Prob}(D_i \geq \varepsilon) \, \mathbf{E}(f(D_i) \mid D_i \geq \varepsilon) \\ &\stackrel{(c)}{\geq} r(n) N_0 \, (1 - \varepsilon^2) \, \mathbf{E}(f(D_i) \mid D_i \geq \varepsilon) \\ &\stackrel{(d)}{\geq} r(n) N_0 \left(\frac{c_3 \, \mathbf{E} D_i}{\sqrt{n \log n}^{\alpha-1}}\right), \end{aligned} \quad (23)$$

where (a) is from (22), the fact that $f(L_i)$ is an increasing function and from (18). Inequality (b) comes from the conditional expectation. Nodes are distributed uniformly on $S^2$, and hence the probability that the distance between a source destination pair exceeds $\varepsilon$, is lower bounded by $(1 - \varepsilon^2)$, which results in (c). When $L_i$ exceeds a constant $\varepsilon$, the $c_2 L_i \sqrt{n \log n}$ term in the denominator of $f(L_i)$ dominates, resulting in (d).

Recollect that $D_i$ is the physical distance on $S^2$ between the source $i$ and destination. Since $\mathbf{E} D_i$ is a constant, (23) results in

$$r(n) \leq c_4 P_0 (n \log n)^{\frac{\alpha-1}{2}} \quad (24)$$

with probability exceeding $1 - \frac{1}{\log n} - 50 \frac{\log n}{n}$. Thus, *w.h.p.*, for sufficiently large $c_5$,

$$\lim_{n \to \infty} \text{Prob}\left( r(n) = c_5 (n \log n)^{\frac{\alpha-1}{2}} \text{ is feasible} \right) = 0. \quad (25)$$

This proves the upper bound, $r(n) = O\left((n \log n)^{\frac{\alpha-1}{2}}\right)$. The upper bound was achieved by analyzing the UWB physical model. The large bandwidth was exploited to demonstrate the negligible effect of interference. The allowed power and rate adaptation were analyzed to develop an optimal power constrained routing strategy, resulting in an upper bound on the uniform throughput capacity. As noted earlier, it is the combination of these mechanisms that allows for the demonstrated bound, an increasing function of $n$.

*D. Area Scaling*

The area of the network has been normalized to unity in the analysis above. Thus, the node density increases linearly with $n$. Consider an alternate scenario where the area of the network $A$ increases with $n$, as $nA_0$. This could represent a situation such as smart homes, where the node density $A_0$ could be a constant. This results in a scaling of all distances by $\sqrt{nA_0}$. The upper bound under this area scaling may be obtained by following the arguments of the previous sections.

The probabilistic arguments for routing optimality and the number of Voronoi cells intersecting a route remain the same. These arguments are independent of the absolute distances. The distance scaling, however, affects the relationship between power and capacity. Following the arguments for the upper bound, we note the point of departure is that, under the new scaling, $\mathbf{E} L_i = \Theta(\sqrt{n})$, and so (24) must be modified to

$$R(n) \leq C_5 P_0 \frac{(\log n)^{\frac{\alpha-1}{2}}}{\sqrt{n}}, \quad (26)$$

where $R(n)$ is the uniform throughput capacity under the new area scaling, with a corresponding modification in (25).

VI. LOWER BOUND ON THROUGHPUT CAPACITY

To provide a lower bound on the capacity, the techniques reviewed in Section II will be useful. The MAC scheme is again chosen as the CDMA MAC, since that was shown to be optimal in Section IV.

We need to demonstrate a feasible routing scheme to provide the lower bound. The routing scheme chosen is the same as in [1]. Thus, as reviewed in Section II, a route is selected for each source-destination pair by following the minimum distance path (segments of great circles), as closely as possible. For such a routing scheme, the number of routes intersecting any cell maybe bounded *w.h.p.*, similar to [1]. Thus, the traffic to be carried by a cell may be upper bounded *w.h.p.* as (8), reproduced below for convenience.

$$\sup_{V \epsilon \mathcal{V}_n} \text{ (Traffic carried by cell V)}$$
$$\leq k_4 r(n) \sqrt{n \log n}. \quad (27)$$

Traffic is relayed from cell to cell till it reaches the cell of the destination node. However, each relay node has a limit on its available capacity. *This limit arises due to the power constraint of the node, unlike [1], where the capacity limit arose from the bandwidth constraint of the network.* From the Voronoi tessellation, we know that every cell is contained in a disk of radius $2\rho$, and so the length of each hop, to reach the next relay node, is at most $8\rho$. Thus, from (14), the relay node has a total capacity $r_i$ bounded as,

$$r_i \geq \frac{c_6 P_0}{N_0} \sqrt{\frac{n}{\log n}}^\alpha. \quad (28)$$

The trade off between the traffic to be carried (27), obtained from routing requirements, and the available capacity (28), obtained by power constraint provides the lower bound. Thus, from (27) and (28), a uniform throughput $r(n)$ is feasible *w.h.p.* if

$$k_4 r(n) \sqrt{n \log n} \leq \frac{c_6 P_0}{N_0} \frac{\sqrt{n}^\alpha}{\sqrt{\log n}^\alpha}. \quad (29)$$

That is,

$$\text{Prob}\left(r(n) = c_7 \frac{n^{\frac{\alpha-1}{2}}}{(\log n)^{\frac{\alpha+1}{2}}} \text{ is feasible}\right) = 1. \quad (30)$$

This proves the lower bound, $r(n) = \Omega(\frac{n^{(\alpha-1)/2}}{(\log n)^{(\alpha+1)/2}})$.

As a side note, it was assumed that each node can transmit and receive simultaneously. However, this restriction can be easily circumvented, by assuming that each link transmits over only half the bandwidth (chosen randomly). Then, the transmission by a node can be thought of as causing an erasure in its own received signal. Thus, as long as each link is encoded with a rate-$\frac{1}{2}$ erasure correction code [31], the throughput will reduce by a factor of at most two, thus satisfying the same order bounds.

*Area Scaling* : As in the case of the upper bound, we can derive the lower bound under the area scaling $A = nA_0$, for which node density is constant. The available capacity (28) is altered due to the dependence of gain $G_{ij}$ (where $G_{ij}$ is used to represent the gain under the new scaling) on the absolute distance measure. Accounting for the scaling, the uniform throughput is bounded as

$$R(n) \leq \frac{C_7}{\sqrt{n}\sqrt{\log n}^{\frac{\alpha+1}{2}}} \quad (31)$$

Thus, for the case of constant node density, both the upper bound and lower bound are decreasing with $n$. The intuitive reason for this is the explicit capacity-distance relationship (14), where capacity decreases with distance.

## VII. PRACTICAL BANDWIDTH SCALING

The large bandwidth $W = \Theta(n(n^2 log n)^{\frac{\alpha}{2}})$, assumed in Section IV to prove the optimality of CDMA MAC is restrictive. However, practical bandwidth scaling schemes can be developed for the unit area and area scaling cases, that require smaller bandwidth. It only needs to be shown that the *lower* bound is achievable with a smaller bandwidth.

**Unit Area case:** Let the bandwidth scaling be $W = \tilde{W}n$. Allot each node a unique disjoint frequency band of bandwidth $\tilde{W}$, disjoint with the bands of other nodes (i.e., a FDMA MAC). Thus, the capacity of link $X_i \to X_j$ is $\tilde{W} \log(1 + \frac{P_{ij}g_{ij}}{N_0\tilde{W}})$. A bandwidth of $\tilde{W} = O(n^{\frac{\alpha}{2}})$ is sufficient to ensure that the capacity is approximately linear in the received power. Therefore, the capacity under this FDMA MAC approximates the CDMA MAC link capacity (12) with a bandwidth $W = \Omega(n^{\frac{\alpha+1}{2}})$.

**Area scaling case:** When the area is scaled as $A = nA_0$, an efficient bandwidth scaling may be obtained, by choosing a hybrid FDMA/CDMA MAC. Recollect that the area of each Voronoi cell is $100 \log n$ in this case, due to area scaling. Form a graph G', with nodes representing vertices, such that two vertices are connected if the corresponding nodes are within a distance of $c_{11}\sqrt{\log n}$ of each other, for some large constant $c_{11}$. Then, the number of cells that have a node connected to a given node is a constant. Each cell has at most $150 \log n$ nodes. So, *w.h.p.*, the degree of G' is upper bounded as $c_{12} \log n - 1$. Now, consider an FDMA/CDMA scheme where the total available bandwidth $W = c_{12}W_0 \log n$ is partitioned equally into $c_{12} \log n$ disjoint frequency bands (FDMA). One band of width $W_0$ is allotted to each node, such that no two nodes that are connected in G' are allotted the same band. A simple greedy algorithm can achieve such a graph coloring [29]. Thus, the MAC chosen is FDMA locally, while CDMA is used to handle the interference from outside the local region. It needs to be shown that $W_0$ can be chosen so that the interference from nodes using the same frequency band (all of which lie outside the local region) is rendered negligible.

To this end, consider the interference caused to a given (receiver) node, by nodes using the same frequency band. Consider the annulus regions formed by circles of radii $\rho_i = (c_{11} - 10) \, i\sqrt{\log n}, \, i \geq 1$, centered on the receiver node under consideration. The number of Voronoi cells in each annulus can be upper bounded by $c_{13}\frac{\rho_{i+1}^2 - \rho_i^2}{\log n}$. Even with the pessimistic assumption that every cell outside the circle of radius $\rho_1$ has one node interfering with the center node (it cannot be more than one due to the local FDMA), the interference from each annulus caused to the center cell is upper bound by $c_{14}\frac{\rho_{i+1}^2 - \rho_i^2}{\rho_i^\alpha \log n}$.

Since the total number of annuli is $c_{15}\sqrt{\frac{n}{\log n}}$, the total interference at the center node is therefore upper bounded as,

$$\begin{aligned}
\text{Interference} &\leq c_{14} \sum_{i=1}^{c_{15}\sqrt{\frac{n}{\log n}}} \frac{\rho_{i+1}^2 - \rho_i^2}{\rho_i^\alpha \log n} \\
&= \frac{c_{16}}{\sqrt{\log n}^\alpha} \sum_{i=1}^{c_{15}\sqrt{\frac{n}{\log n}}} \frac{2i+1}{i^\alpha} \\
&\leq \frac{c_{16}}{\sqrt{\log n}^\alpha} \sum_{i=1}^{c_{15}\sqrt{\frac{n}{\log n}}} \frac{3}{i^{\alpha-1}} \\
&\leq \frac{c_{16}}{\log n} \sum_{i=1}^{c_{15}\sqrt{\frac{n}{\log n}}} \frac{3}{i} \quad \text{for any } \alpha \geq 2 \\
&\stackrel{(a)}{\leq} c_{17}. 
\end{aligned} \quad (32)$$

where (a) arises from the bound $\sum_{i=1}^{y} \frac{1}{i} \leq 1 + \log(y)$. Thus, a sufficiently large constant per-node bandwidth $W_0$ is sufficient to render interference negligible with respect to noise $N_0 W_0$, if $\alpha \geq 2$. Since there are $\Theta(\log n)$ frequency bands required, the total required system bandwidth is $W = \Theta(\log n)$. Intuitively, with area scaling, the closest and (therefore) dominant interferers are moved away, resulting in a smaller interference. However, as demonstrated in Section VI, this area scaling results in a decreasing capacity function. A practical bandwidth scaling for $\alpha < 2$ does not seem obvious.

## VIII. Conclusion

In this paper, the capacity of an UWB (large bandwidth) ad-hoc wireless network with a power constraint was studied. Examples of such a network include UWB and sensor networks. It was shown that for such a network, consisting of $n$ randomly distributed identical nodes over a unit area, with probability approaching one (as $n \to \infty$), the uniform throughput capacity $r(n)$ is $O\left((n \log n)^{\frac{\alpha-1}{2}}\right)$ (upper bound) and $\Omega(\frac{n^{(\alpha-1)/2}}{(\log n)^{(\alpha+1)/2}})$ (lower bound). Thus, the throughput capacity $r(n)$ for such a random ad-hoc network is $\tilde{\Theta}(n^{(\alpha-1)/2})$. Interestingly, this bound demonstrates an increasing per-node throughput, in comparison to the decreasing per-node throughput shown in [1]. The key reason for this dramatically contrasting result is that the assumed UWB physical model is characterized by finite power, large bandwidth, and the explicit use of link adaptation. These characteristics combine to alleviate interference and exploit node density effectively while delivering traffic. Thus, the properties of the physical layer dramatically alter the ad hoc network capacity.

Practical bandwidth scaling results were derived to show that the assumption of arbitrarily large bandwidth is not excessively restrictive. Further, the optimal MAC and routing, which can achieve network capacity, were specified - namely a CDMA MAC and a power-constrained routing. The demonstrated results and protocols may be utilized as a guideline for practical UWB network design.

## IX. Appendix

The ideas of Vapnik-Chervonenkis (VC) theory are reviewed in this section. A finite set of points $X$ (such as nodes) of size $n$ and a set of subsets $\mathcal{F}$ (such as Voronoi cells) is considered. $X$ is said to be shattered by $\mathcal{F}$ if for every subset $B$ of $X$ there is a set $F \epsilon \mathcal{F}$ such that $X \bigcap F = B$. $\mathcal{F}$ generates all subsets of $X$. The VC-dimension of $\mathcal{F}$ is defined as the supremum of the sizes of all finite sets that can be shattered by $\mathcal{F}$ [36], [37].

An underlying probability distribution ($PD$) is assumed. An i.i.d sequence $X = X_1 ... X_n$ is chosen with distribution $PD$. The relative frequencies of events are $N(F) = \frac{1}{n} \sum_{i=1}^{n} I(X_i \epsilon F)$. Sets of finite VC-dimension obey a uniform convergence in the weak law of large numbers, i.e., relative frequencies of the events converge to their probabilities uniformly. Formally,

*VC THEOREM:* If $\mathcal{F}$ is a set of finite VC-dimension, and $X_i$ is a sequence of i.i.d. random variables with common probability distribution $PD$, then for every $\varepsilon, \delta > 0$,

$$\text{Prob}\left(|N(F) - PD(F)| \leq \varepsilon \ \forall F \ \epsilon \ \mathcal{F}\right) > 1 - \delta, \quad (33)$$

for sufficiently large n. The VC-theorem (33) is a stronger statement than the weak law of large numbers due to the uniformity over all events.

The proof [37] is developed by first defining the growth function $m(n)$, which is the maximum number of subsets of an $n$ sized sample $X$ generated by the set of events $\mathcal{F}$. It is then proved, for sets with finite VC-dimension $vc$, that the growth function is upper bounded as

$$m(n) \leq n^{vc} + 1 \quad s.t. \ n \geq vc. \quad (34)$$

The probability that the relative frequency of a *particular event* $F$ exceeds its expected value by $\varepsilon$ is exponentially decreasing in the sample size $n$. To bound the probability over the class of events $\mathcal{F}$ a union bound is applied. However the union is now taken over $m(n)$ events since only $m(n)$ events are distinct. Thus

$$\text{Prob}\left(\sup_{F \epsilon \mathcal{F}} |N(F) - PD(F)| \geq \varepsilon\right) \leq m(n) * e^{-n\varepsilon^2}. \quad (35)$$

The finite VC-dimension implies m(n) grows slower that the exponential in $n$, and hence the weak law holds uniformly.

## References

[1] P. Gupta and P. R. Kumar, "Capacity of wireless networks," *IEEE Transactions on Information Theory*, vol. 46(2), pp. 338-404, March 2000.

[2] M. Z. Win and R. A. Scholtz, "On the robustness of ultra wide bandwidth signals in dense multipath environment," *IEEE Comm Letters*, vol. 2(2), pp. 51-53, Feb. 1998.

[3] R. Ramanathan and J. Redi, " A brief overview of ad hoc networks: challenges and directions," *IEEE Communications Magazine*, vol. 40(5), pp. 20-22, May 2002.

[4] H. Luo, S. Lu, and V. Bharghavan, "A new model for packet scheduling in multihop wireless networks," *ACM MobiCom '00*, pp. 76-86, Aug. 2000.

[5] X. L. Huang and B. Bensaou, "On max-min fairness and scheduling in wireless ad-hoc networks: analytical framework and implementation," *Proc. MobiHOC '00*, pp. 221-231, 2001.

[6] N. Bambos, S. Chen, and G. Pottie, "Radio link admission algorithms for wireless networks with power control and active link quality protection," *Proc. IEEE Infocom '95*, vol. 1, pp. 97-104, March 1996.

[7] Dapeng Wu, and R. Negi, "Effective capacity: A wireless channel model for support of quality of service," *Proc. IEEE Globecom*, vol. 1, pp. 695-699, San Antonio, 2001.

[8] C. E. Perkins and E. M. Royer, "Ad-hoc on-demand distance vector routing," *Proceedings of the 2nd IEEE Workshop on Mobile Computing Systems and Applications*, pp. 90-100, 1999.

[9] E. M. Royer and C. K, Toh, "A review of current routing protocols for ad hoc networks," *IEEE Personal Communications*, vol. 6, pp 46-55, April 1999.

[10] D. B. Johnson and D. Maltz, "Dynamic source routing in ad hoc wireless networks," *Mobile Computing*, vol. 353, Kluwer Academic Publishers, 1996.

[11] G. Kim, A. Rajeswaran and R. Negi, "Joint power adaptation, scheduling and routing framework for wireless ad hoc networks,"*Proc. IEEE SPAWC '05*, pp. 745-749, June 2005.

[12] M. Grossglauser and D. Tse, "Mobility increases the capacity of ad-hoc wireless networks," *Proceedings of IEEE Infocom 2001*, pp. 1360-1369, April 2001.

[13] S. Toumpis, "Capacity bounds for three classes of wireless networks: Asymmetric, cluster and hybrid," *Proc. of ACM Mobihoc 2004*, pp. 133-144, May 2004.

[14] N. Bansal and Z. Liu, "Capacity, delay and mobility in wireless ad-hoc networks," *Proceedings of IEEE Infocom 2003*, vol. 2, pp. 1553-1563, April 2003.

[15] A. El Gamal, J. Mammen, B. Prabhakar, and D. Shah, "Throughput-delay trade-off in wireless networks," *Proc. of IEEE Infocom, 2004*, vol. 1, pp. 464-475, March 2004.

[16] J. Li, C. Blake, D. D. Couto, H. Lee, and R. Morris, "Capacity of ad hoc wireless networks," *ACM MobiCom 2001*, pp. 61-69, July 2001.

[17] S. Yi, Y. Pei, and S. Kalyanaraman, "On the capacity improvement of ad hoc wireless networks using directional antennas," *Proc. MobiHOC*, pp. 108-116, June 2003.

[18] S. Toumpis and A. J. Goldsmith, "Large wireless networks under fading, mobility and delay constraints," *IEEE Infocom 2004*, vol. 1, pp. 609-619, March 2004.

[19] L. Xie and P. R. Kumar, "A network information theory for wireless communication: Scaling laws and optimal operation," *IEEE Transactions on Information Theory*, vol. 50(5), pp. 748-767, May 2004.

[20] A. Rajeswaran, V. S. Somayazulu, J. R. Foerster, "Rake performance for a pulse based UWB system in a realistic UWB indoor channel," *Proceedings of IEEE Int. Conference on Communications (ICC)*, pp. 2879-2883, May 2003.

[21] D. Cassioli, M. Z. Win, F. Vatalaro, A. Molisch, "Performance of low-complexity rake reception in a realistic UWB channel," *Proceedings of IEEE Int. Conference on Communications (ICC)*, pp. 763-767, May 2002.

[22] J. Foerster, E. Green, S. Somayazulu, and D. Leeper, "Ultra-wide band technology for short or medium range wireless communications," *Intel Technology Journal*, vol. 5(2). Available at http://developer.intel.com/technology/itj/q22001/

[23] IEEE 802.15 WPAN High Rate Alternative PHY Task Group 3a, accessible at http://grouper.ieee.org/groups/802/15/pub/TG3a.html

[24] R. Fontana, A. Ameti, E. Richley, L. Beard, D. Guy, "Recent advances in ultra wideband communications systems," *Digest of IEEE Conference on Ultra Wideband Systems and Technologies*, pp. 129-133, May 2002.

[25] R. Negi and A. Rajeswaran, "Capacity of power constrained ad-hoc networks," *INFOCOM '04*, vol. 1, pp. 443-453, March 2004.

[26] H. Zhang and J. Hou, "Capacity of wireless ad hoc networks under ultra wide band with power constraint," *Proc. of IEEE Infocom, 2005*, March 2005.

[27] M. Franceschetti, O. Dousse, D. Tse, and P. Thiran, "Closing the gap in the capacity of random wireless networks," *Proc. ISIT 04*, pp. 439, June 2004.

[28] J. G. Proakis, *Digital Communications, 3rd Ed.*, McGraw-Hill, 1995.

[29] J. A. Bondy and U. Murthy, *Graph Theory with Applications*, New York, Elsevier, 1976.

[30] I. Akyildiz, S. Weilian, Y. Sankarasubramaniam and E. Cayirci, "A survey on sensor networks," *IEEE Communications Magazine*, vol. 40, issue 8, pp. 102-114, Aug. 2002.

[31] T. M. Cover and J. A. Thomas, *Elements of Information Theory*, John Wiley, 1991.

[32] R. Negi and J. Cioffi, "Delay-constrained capacity with causal feedback," *IEEE Transactions on Information Theory*, vol. 48, pp. 2478-2494, Sept. 2002.

[33] R. Negi and J. Cioffi, "Stationary schemes for optimal transmission over fading channels with delay constraint," *Proc. IEEE VTC '00*, pp. 358-361, Sep. 2000.

[34] C. Peraki and S. Servetto, "On the maximum stable throughput problem in random wireless networks with directional antennas," *ACM Mobihoc '03*, pp. 76-87, March 2003.

[35] D. Bertsekas and R. Gallager, *Data Networks*, Prentice Hall, 1992.

[36] V. N. Vapnik, *Statistical Learning Theory*, John Wiley & sons, 1998.

[37] V. N. Vapnik and A. Chervonenkis, "On the uniform convergence of relative frequencies of events to their probabilities," *Theory of Probab. and its Applications*, vol. 16(2), pp. 264 - 280, 1971.

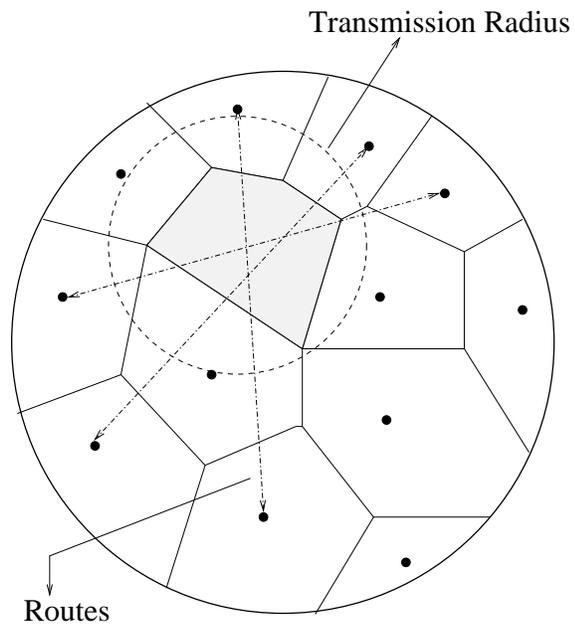

Fig. 1. Voronoi Tessellation

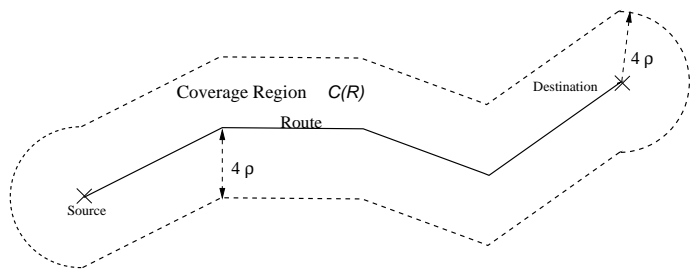

Fig. 2. Coverage region of a Route